\documentclass[twocolumn,showpacs,preprintnumbers,amsmath,amssymb,prb]{revtex4}

\usepackage{graphicx}% Include figure files
\usepackage{bm}% bold math

\begin{document}
\draft
\title{Anomalous heat conduction in a carbon nanowire: Molecular dynamics calculations}

\author{Gang Wu}
\email{wugaxp@gmail.com}
\author {Jinming Dong}
\email{jdong@nju.edu.cn}

\affiliation{Group of Computational Condensed Matter Physics,
National Laboratory of Solid State Microstructures and Department
of Physics, Nanjing University, Nanjing 210093, P. R. China}

\begin{abstract}

Heat conduction of a real quasi-one dimensional material, the
finite length carbon nanowire (CNW), inserted into the
single-walled carbon nanotube (SWNT) has been studied by the
molecular dynamical (MD) method, in which both of the longitudinal
as well as transverse motions of the chain atoms in the SWNT have
been permitted. It is found that the thermal conductivity $\kappa
$ of the carbon nanowire is very high at room temperature, and
diverges more likely with the chain length logarithmically.

\end{abstract}


\date{\today}
\maketitle

%\begin{multicols}{2}

\section{Introduction }

When a small temperature gradient exists between two boundaries of
a material, it is expected that heat will be transported through
the material, which usually obeys the Fourier's law of conduction
($\mathord{\buildrel{\lower3pt\hbox{$\scriptscriptstyle\rightharpoonup$}}\over
{j}} = - \kappa \nabla T)$, a well-known fact in three-dimensional
systems, where
$\mathord{\buildrel{\lower3pt\hbox{$\scriptscriptstyle\rightharpoonup$}}\over
{j}} $ is the heat amount flowing through a unit surface per unit
time,$\nabla T$ is the gradient of the temperature field over the
material, and $\kappa $ is defined as the thermal conductivity.
However, it is not clear whether the Fourier's law is still
correct in the lower (one or two) dimensional systems, which
stimulated a great interest in the past several years. It has been
shown [1-14] that heat conduction exhibits anomalous behavior in
some one dimensional systems. For example, in the one-dimensional
(1D) integrable systems, such as the harmonic chain and the
monoatomic Toda model, no temperature gradient can be set up. In
some 1D nonintegrable systems, such as the F-K model [2], the
discrete $\phi ^4$ model [4] and the Lorentz model [5], the
temperature gradient is uniform, and the heat conductivity $\kappa
$ is a constant, being independent of system size, which means
these 1D systems still obey Fourier's law. However, in some other
anharmonic 1D systems, like the Fermi-Pasta-Ulam (FPU) $\beta
$-model [6, 2], the diatomic Toda chain [7], or in 1D
hard-particle gases with alternating masses [7, 8, 9], the
temperature gradient can be formed as $\frac{dT}{dx} \sim L^{ -
1}$, and their $\kappa $ scales as $\kappa \sim L^\alpha $, where
$L$ is the system size, and $\alpha > 0$. Recently, Wang
\textit{et al.} [14] studied the anomalous thermal conduction in
1D polymer chains with a modeled Hamiltonian, and found three
types of divergent exponent $\alpha $ in them, which are caused by
different couplings between longitudinal and transverse motions.
However, although a great deal of theoretical researches on the
problem had been made in the past several years, there exists
right now still a lot of controversies about the divergence
behavior of the thermal conductance in low dimensional systems,
and many important and fundamental questions in this field remain
unsolved.

But all these systems said above seem to be far from real physical
materials. What will happen for the thermal conduction in a real
low dimensional material? Does it follow the Fourier's law or not?
All of these problems not only stimulate fundamental research
interests, but also have great potential applications in the
thermal conduction of the nano-materials. When the dimensions of
electronic devices shrink to nano-scale due to the fast progress
in the present microelectronic technology, the thermal conduction
problem becomes more and more important because a significant
energy should be dissipated in a much smaller compact space. And
the divergence of the thermal conductance with the length in the
low dimensional materials promises possibility of making the more
outstanding heat dissipation nano-materials, solving the thermal
dissipation problem coming from the miniaturization of the
electronic and optical devices. So, it is very interesting to
study the heat conduction in a real 1D or quasi-1D physical
system.

Recently, carbon nanotubes (CNTs) have attracted much attention
due to their remarkable electronic, thermal and mechanical
properties [15]. The diameter of a typical nanotube ranges from
several to several tens angstroms, and that of the smallest one is
only $3$ \AA[16]. While their lengths can be several $\mu $m, and
even reach to several mm, being much larger than their diameters.
So, carbon nanotube can be thought as a very well 1D system. In
fact, many experiments and numerical simulations have found that
the thermal conductivity $\kappa $ of SWNTs is extremely high
although there exists a distribution of the obtained $\kappa $
values. For example, the observed room-temperature thermal
conductivity of SWNT rope is about 1750$\sim $5800 W/mK [17], and
for individual multiwalled carbon nanotubes (MWNTs), this value is
larger than 3000 W/mK [18]. Using equilibrium and nonequilibrium
molecular dynamics simulations, recent numerical simulations also
give similar results. Berber \textit{et al.} [19] found that, for
an isolated (10, 10) nanotube at room temperature, $\kappa \approx
6600$ W/mK. S. Maruyama \textit{et al.} [20] claimed that $\kappa
$ of (5, 5) nanotube diverges as a power law relation with the
tube length, and got a rather smaller $\kappa $ value of about 150
$\sim $ 500 W/mK for the (5, 5) tube. Zhang and Li [21] studied
three armchair SWNTs, i.e., (5, 5), (10, 10) and (15, 15), and
found that their $\kappa's$ diverge as a power law, too, with
their $\kappa $ values of about 700 $\sim $ 2200 W/mK, higher than
that in Ref.[20]. Yao \textit{et al.} [22] also studied the same
three armchair tubes, and found their $\kappa's$ could diverge
with their lengths, and have the same higher $\kappa $ value of
about 400 $\sim $ 2500 W/mK.

At the same time, a new type of carbon structure, carbon nanowires
(CNWs) [23] have been discovered in the cathode deposits prepared
by hydrogen arc discharge evaporation of carbon rods. The CNWs are
made of extraordinarily long 1D linear carbon chains consisting of
more than 100 carbon atoms inserted into the innermost tube (7 \AA
diameter) of MWNTs. The CNW can be considered as another good
example of the real 1D physical system. In this paper, we will
investigate in detail the heat conduction in the CNWs and pay much
our attention to the divergence behavior of its thermal
conductivity.

In what follows we firstly introduce the model Hamiltonian and calculation
method. Then in Sec. III we give the main numerical results and discuss the
divergence of thermal conductivity in the CNWs. The paper ends with some
concluding remarks in Sec. IV.

\section{Model}

As well known, a bare long carbon chain can not be stable, and so
we suppose a chain of N carbon atoms with the same mass $m_c $ is
inserted into a single-walled carbon nanotube. The interaction
between chain atoms is simulated by the Tersoff-Brenner bond order
potential [24], and the interaction between carbon chain and
outside nanotube is described by Lennard-Jones(LJ) potential,

\begin{equation}
\label{eq1}
u\left( x \right) = 4\varepsilon \left[ { - \left( {\frac{\sigma }{x}}
\right)^6 + \left( {\frac{\sigma }{x}} \right)^{12}} \right].
\end{equation}

In our simulation, $\varepsilon $ and $\sigma $ are taken as 2.41
mev and 3.4 \AA [25], respectively. Then the Hamiltonian of the
chain system can be written as

\begin{equation}
\label{eq2}
H = \sum\limits_{i = 1}^N
{\frac{\mathord{\buildrel{\lower3pt\hbox{$\scriptscriptstyle\rightharpoonup$}}\over
{p}} _i^2 }{2m_i }} + \textstyle{1 \over 2}\sum\limits_{i,j = 1}^N {V_{ij} +
U_i } ,
\end{equation}

\noindent
where

\begin{equation}
\label{eq3}
V_{ij} = f_c \left(
{\mathord{\buildrel{\lower3pt\hbox{$\scriptscriptstyle\rightharpoonup$}}\over
{r}} _{ij} } \right)\left[ {a_{ij} f_R \left(
{\mathord{\buildrel{\lower3pt\hbox{$\scriptscriptstyle\rightharpoonup$}}\over
{r}} _{ij} } \right) + b_{ij} f_A \left(
{\mathord{\buildrel{\lower3pt\hbox{$\scriptscriptstyle\rightharpoonup$}}\over
{r}} _{ij} } \right)} \right].
\end{equation}

Here, $f_c \left(
{\mathord{\buildrel{\lower3pt\hbox{$\scriptscriptstyle\rightharpoonup$}}\over
{r}} _{ij} } \right)$ is a cut-off function, $f_R \left(
{\mathord{\buildrel{\lower3pt\hbox{$\scriptscriptstyle\rightharpoonup$}}\over
{r}} _{ij} } \right)$ and $f_A \left(
{\mathord{\buildrel{\lower3pt\hbox{$\scriptscriptstyle\rightharpoonup$}}\over
{r}} _{ij} } \right)$ are two Morse type functions which represent
the attractive and repulsive effects of the potential,
respectively, and $a_{ij} $ and $b_{ij} $ are two parameters
describing bond order and bond angular effects. Full details of
the model potential are available in the original paper of Tersoff
and Brenner [24]. $U_i $ is external potential exerted by outside
nanotube. $m_i $ is the mass of chain atoms, $i.e.$, the carbon
atom mass. For simplicity, we assume the carbon atoms on outside
nanotube is distributed \textit{continuously}, which is well known
as the continuum model and used in a lot of systems [26-37]. For
example, based upon the same idea, L.A. Girifalco \textit{et al.}
developed a simple universal graphitic potential in their paper
[37]. Now, following them, we take external potential as:

\begin{equation}
\label{eq4}
U\left( r \right) = n_\sigma \int {u\left( x \right)} d\Sigma ,
\end{equation}

\noindent
where $r$ and $x$ represent the time-dependent distances of the wire atom to
tube axis and surface element $d\Sigma $, respectively. $n_\sigma $ is the
mean surface density of tube atoms. In the cylindrical coordinates, $U\left(
r \right)$ can be expressed as:

\begin{equation}
\label{eq5}
U\left( r \right) = n_\sigma \int {u\left( x \right)} \rho d\theta dz,
\end{equation}

\noindent
where

\begin{equation}
\label{eq6}
x = \sqrt {\left( {\rho \cos \theta - r} \right)^2 + \rho ^2\sin ^2\theta +
z^2} ,
\end{equation}

\noindent
and $\rho $ is the radius of outside tube, $\theta $ and $z$ is another two
cylinder coordinates of $d\Sigma $.

Thus, when $r \ne 0$, the surface integral can be simplified to give

\begin{equation}
\label{eq7}
U\left( r \right) = 3\pi \rho n_\sigma \varepsilon \left[ { - \frac{\sigma
^6}{\left( {4\rho r} \right)^{\textstyle{5 \over 2}}}I_5 +
\frac{21}{32}\frac{\sigma ^{12}}{\left( {4\rho r} \right)^{\textstyle{{11}
\over 2}}}I_{11} } \right],
\end{equation}

\noindent
where

\begin{equation}
\label{eq8}
I_n = \int_0^{\textstyle{\pi \over 2}} {\frac{dt}{\left( {a^2 + \sin ^2t}
\right)^{\textstyle{n \over 2}}}} ,
\end{equation}

\[
a^2 = \frac{\left( {\rho - r} \right)^2}{4\rho r}.
\]

Here, $I_n $ is an integral related to the hypergeometric function, which
can be made exactly in an expanded series, and obtained result is expressed
as

\[
I_n = \textstyle{\pi \over 2}b^{ - n}\left[ {1 + \sum\limits_{m =
1}^\infty {\frac{\left( {2m - 1} \right)!!\left( {2m + n - 2}
\right)!!}{\left( {n - 2} \right)!!\left[ {\left( {2m} \right)!!}
\right]^2b^{2m}}} } \right],
\]

\begin{equation}
\label{eq9}
b^2 = a^2 + 1 = \frac{\left( {\rho + r} \right)^2}{4\rho r}.
\end{equation}

But when $b \to 1$, $i. e.$, $r \sim \rho $, the summation in the $I_n $ will
converge very slowly. So, in that case, after taking some algebra technique,
a more efficient expression can be finally obtained:

\begin{widetext}
\begin{equation}
\label{eq10}
I_{2k + 1} \approx \left\{ {\frac{1}{\left( {2k - 1} \right)!!}\left(
{\frac{2}{a^2}} \right)^k\sum\limits_{m = 0}^{k - 1} {\frac{\left[ {\left(
{2m} \right)!} \right]^2}{\left[ {m!} \right]^3}\frac{\left( {k - m - 1}
\right)!}{2}\left( {\frac{a}{4}} \right)^{2m}} + \frac{\left( {2k - 1}
\right)!!}{\left( {2k} \right)!!}} \right\}.
\end{equation}
\end{widetext}

Although this expression is an approximate one, but when $a$ is
small enough, it can give very accurate result of $I_n $, and
needs only to take a few terms in its summation. Thus combining
Eq. 7, 8, 9 with Eq. 10, we obtain an efficient external
potential, representing in high precision the interaction between
the chain atoms and the outside tube.

In this work, we only consider armchair SWNT (5, 5) as the outside
tube because its radius is about 3.4 \AA, which is the closest to
that of the innermost tube observed experimentally [23]. And the
average equilibrium distance between the chain atoms is set to be
$a$ = 1.84 \AA, which means there are four carbon atoms in three
periods of outside armchair nanotube. We should mention that at
present there are \textbf{NO experimental data} about the distance
between carbon atoms in the nanowire, which is so selected from
the \textbf{commensurability} between both periods of nanowire and
outside SWNT. In fact, we can choose different $a$ values to study
its effect on the thermal conduction of the CNW, which is beyond
scope of the present investigation and will be left for future
study.

We determine the heat current in a temperature gradient by
nonequilibrium molecular dynamics method. Two atoms at each end of
the CNW are subject to heat baths at $T_L $ and $T_H $
respectively, which usually can be simulated by Nos\'{e}-Hoover
thermostats [38]. The equations of motion for these four atoms are
written as

\[
\ddot
{\mathord{\buildrel{\lower3pt\hbox{$\scriptscriptstyle\rightharpoonup$}}\over
{r}} }_i = - \zeta _L \dot
{\mathord{\buildrel{\lower3pt\hbox{$\scriptscriptstyle\rightharpoonup$}}\over
{r}} }_i +
\mathord{\buildrel{\lower3pt\hbox{$\scriptscriptstyle\rightharpoonup$}}\over
{f}} _i ,
\]

\begin{equation}
\label{eq11}
\ddot
{\mathord{\buildrel{\lower3pt\hbox{$\scriptscriptstyle\rightharpoonup$}}\over
{r}} }_j = - \zeta _R \dot
{\mathord{\buildrel{\lower3pt\hbox{$\scriptscriptstyle\rightharpoonup$}}\over
{r}} }_j +
\mathord{\buildrel{\lower3pt\hbox{$\scriptscriptstyle\rightharpoonup$}}\over
{f}} _j ,
\end{equation}

\noindent
where $f_i $ is the force applied on the $i$th carbon atom. The thermal
variables $\zeta _L $ and $\zeta _R $ evolve according to the equations

\begin{equation}
\label{eq12}
\dot {\zeta }_{L,R} = \frac{1}{Q}\left( {\sum\limits_i
{\frac{\mathord{\buildrel{\lower3pt\hbox{$\scriptscriptstyle\rightharpoonup$}}\over
{p}} _i^2 }{m_i } - gk_B T} } \right),
\end{equation}

\[
Q = gk_B T\tau ^2.
\]

The number of degrees of freedom for particles in thermostats is given by $g
= 6$, and $\tau $ is the relaxation time of the heat bath. The rest of the
atoms follows the equations of motion

\begin{equation}
\label{eq13}
\ddot
{\mathord{\buildrel{\lower3pt\hbox{$\scriptscriptstyle\rightharpoonup$}}\over
{r}} }_i =
\mathord{\buildrel{\lower3pt\hbox{$\scriptscriptstyle\rightharpoonup$}}\over
{f}} _i ,
\quad
i = 3, \cdots ,N - 2,
\end{equation}

\noindent
and fixed boundary conditions are assumed for the zeroth and (N+1)th atoms
($\mathord{\buildrel{\lower3pt\hbox{$\scriptscriptstyle\rightharpoonup$}}\over
{r}} _0 \equiv \left( {0,0,0} \right)$,
$\mathord{\buildrel{\lower3pt\hbox{$\scriptscriptstyle\rightharpoonup$}}\over
{r}} _{N + 1} \equiv \left( {0,0,(N + 1)a} \right))$.

We first use an eighth-order Runge-Kutta algorithm to solve the
ordinary differential equations, which provides more accurate
results than those of the usual fourth-order Runge-Kutta
algorithm. The time step is chosen from $h = 0.01$ to $0.05$ in
the unit of 0.035267 ps. Typical total MD steps are taken as
$10^7$ to $10^8$. And for comparison, we also use the velocity
Verlet algorithm [39] in the same evolution.

The total heat flux can be expressed as follows:

\begin{equation}
\label{eq14}
\mathord{\buildrel{\lower3pt\hbox{$\scriptscriptstyle\rightharpoonup$}}\over
{J}} \left( t \right) = \frac{d}{dt}\sum\limits_i
{\mathord{\buildrel{\lower3pt\hbox{$\scriptscriptstyle\rightharpoonup$}}\over
{r}} _i \left( t \right)\varepsilon _i \left( t \right)} ,
\end{equation}

\noindent
where
$\mathord{\buildrel{\lower3pt\hbox{$\scriptscriptstyle\rightharpoonup$}}\over
{r}} _i \left( t \right)$is the time-dependent coordinate of the wire atom
$i$, and $\varepsilon _i \left( t \right)$ is its total energy, which
contains both of the kinetic and the potential energies. And after
introducing the force on the atom $i$ from atom $j$, i.e.,
$\mathord{\buildrel{\lower3pt\hbox{$\scriptscriptstyle\rightharpoonup$}}\over
{F}} _{ij} = \frac{\partial \varepsilon _i }{\partial
\mathord{\buildrel{\lower3pt\hbox{$\scriptscriptstyle\rightharpoonup$}}\over
{r}} _j }$, the instantaneous local heat current per particle can be
expressed as:

\begin{equation}
\label{eq15}
\mathord{\buildrel{\lower3pt\hbox{$\scriptscriptstyle\rightharpoonup$}}\over
{J}} \left( t \right) = \sum\limits_i {\dot
{\mathord{\buildrel{\lower3pt\hbox{$\scriptscriptstyle\rightharpoonup$}}\over
{r}} }_i \varepsilon _i + \sum\limits_{i,j,i \ne j}
{\mathord{\buildrel{\lower3pt\hbox{$\scriptscriptstyle\rightharpoonup$}}\over
{r}} _{ij} \left(
{\mathord{\buildrel{\lower3pt\hbox{$\scriptscriptstyle\rightharpoonup$}}\over
{F}} _{ij} \cdot \dot
{\mathord{\buildrel{\lower3pt\hbox{$\scriptscriptstyle\rightharpoonup$}}\over
{r}} }_i } \right)} }
\end{equation}

\noindent
where
$\mathord{\buildrel{\lower3pt\hbox{$\scriptscriptstyle\rightharpoonup$}}\over
{r}} _{ij} =
\mathord{\buildrel{\lower3pt\hbox{$\scriptscriptstyle\rightharpoonup$}}\over
{r}} _i -
\mathord{\buildrel{\lower3pt\hbox{$\scriptscriptstyle\rightharpoonup$}}\over
{r}} _j $ is the relative distance between atom $i$ and $j$.

Because of the linear temperature distribution, the classical definition on
the heat conductivity can be used, leading to:

\begin{equation}
\label{eq16}
\kappa = JN / \left( {T_L - T_R } \right)
\end{equation}

When $N \to \infty $, above expression corresponds to the
coefficient of heat conductivity of the chain under temperature $T
= \left( {T_L + T_R } \right) / 2$. In our MD process, $T_L$ and
$T_R$ are kept as 0.03 and 0.025, which correspond to real 290K
and 348K in practice, respectively.

The alternative way to calculate the heat conductivity of the chain is based
on the Green-Kubo formula [40]

\begin{equation}
\label{eq17}
\kappa _s = \frac{1}{k_B T^2N}\int_0^t {\left\langle {J\left( \tau
\right)J\left( 0 \right)} \right\rangle } d\tau ,
\end{equation}

\noindent
where $J\left( t \right)$ is defined in Eq. 14.

In our calculations, we found that these two definitions always give almost
equal results (the difference between them never exceeds a few percent).

\noindent

\section{numerical result and discussions}

\begin{figure}
\includegraphics[width=\columnwidth]{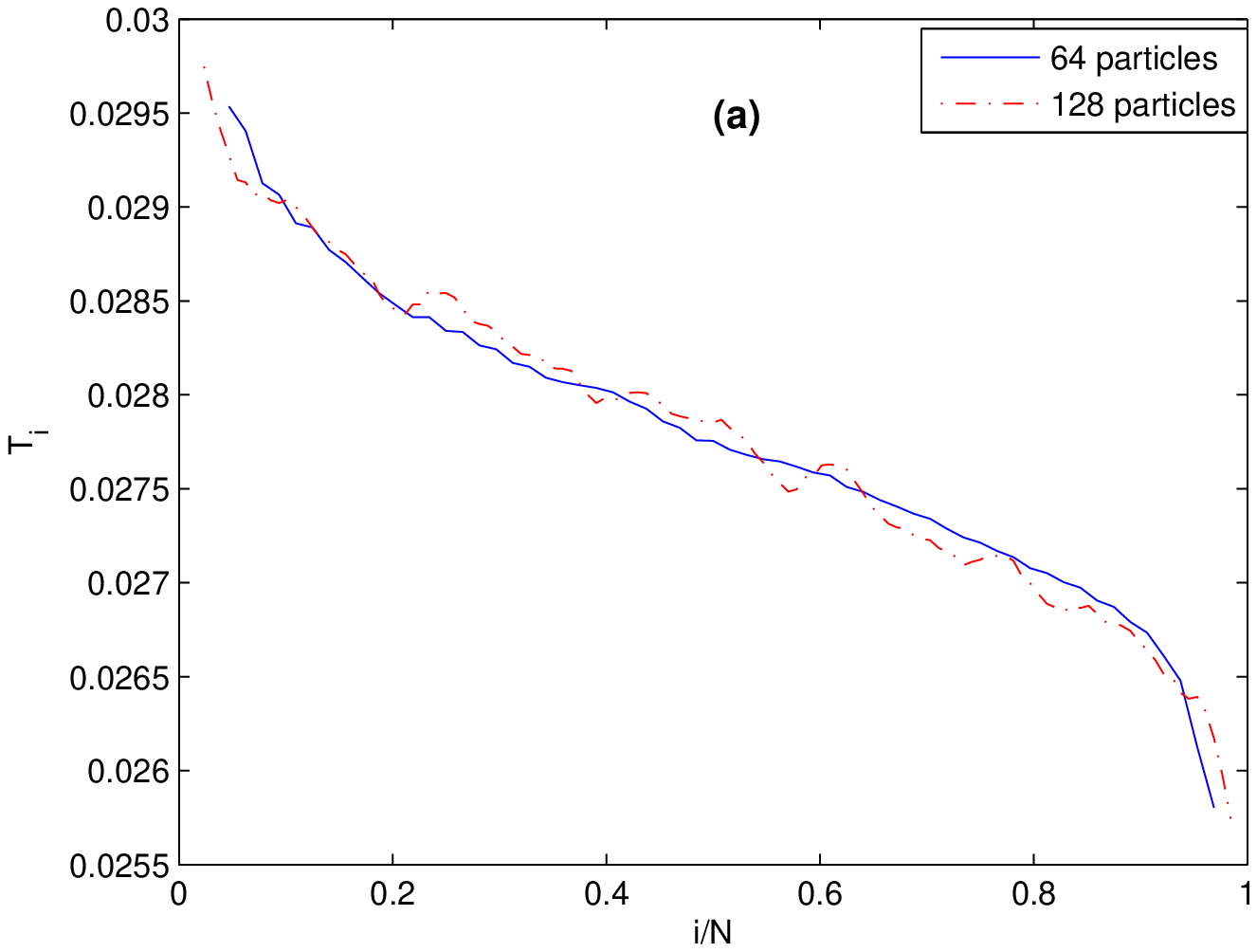}
\label{fig1a}
\includegraphics[width=\columnwidth]{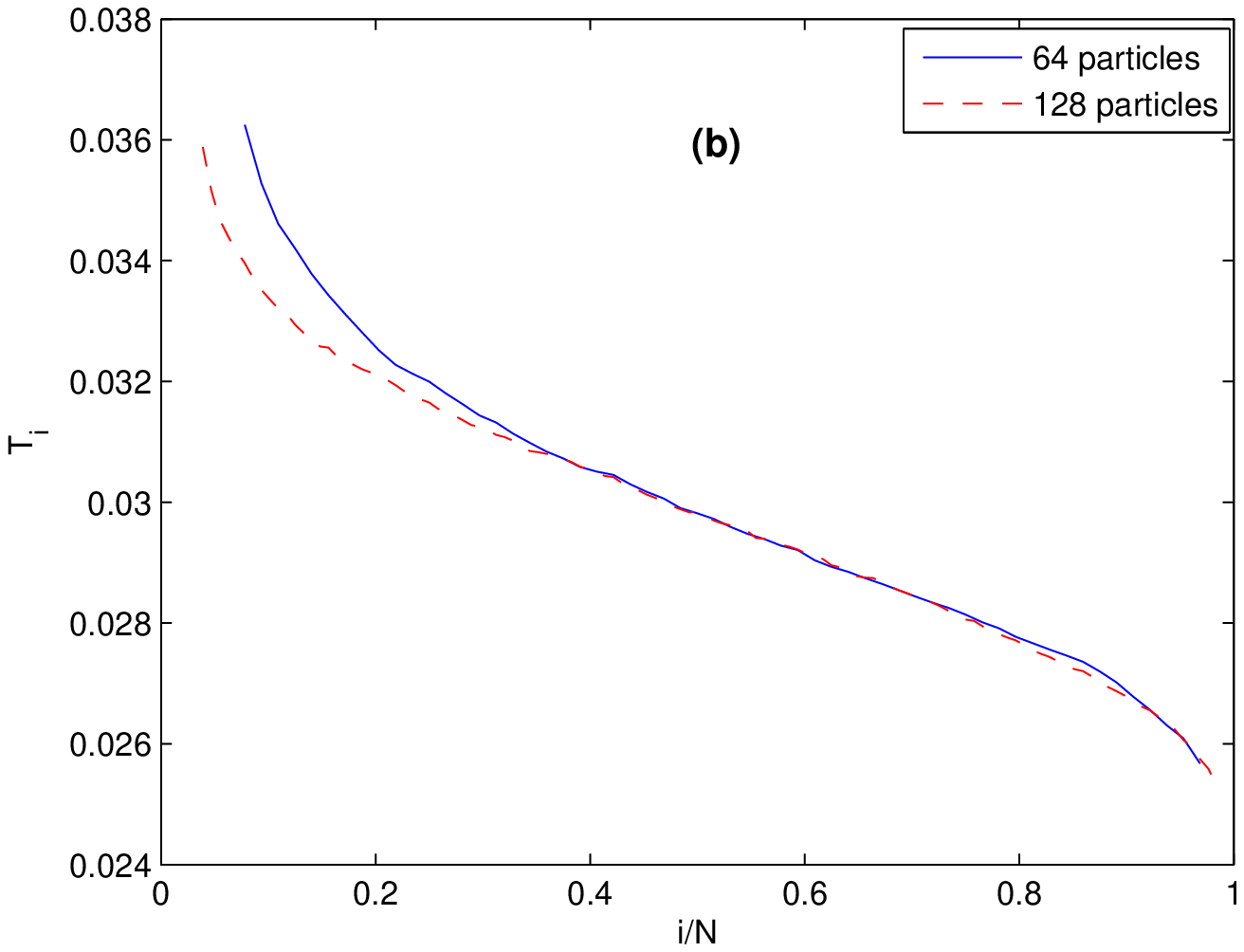}
\label{fig1b} \caption{The temperature profiles on the chain at
$T_L = 0.03$ and $T_R = 0.025$, with $N$ = 64 (solid line) and 128
(dash-dot lines). The averages are carried over a time interval of
$10^4 \sim 10^5$. The distance between CNW atoms is 1.844$\mathop
A\limits^ \circ $. a). Eighth-order Runge-Kutta algorithm. b).
Velocity Verlet algorithm.}
\end{figure}

In Fig. 1 we show the temperature distribution on the chain,
calculated by both eighth-order Runge-Kutta algorithm [Fig. 1(a)]
and velocity Verlet algorithm [Fig. 1(b)]. It is seen from Fig.
1(a) and 1(b) that the linear temperature distribution can be
obtained with both algorithms, but there are still some small
differences between them. The velocity Verlet algorithm is a very
efficient one, by which, a very smooth temperature distribution is
obtained, but in this case, the Nos\'{e}-Hoover boundary condition
on the left end of the chain with higher temperature could not be
well treated unless the chain is long enough, which may result
from the sensitivity of this algorithm to the thermostat boundary
condition. So, we will mainly use the Runge-Kutta algorithm in
this work, and the velocity Verlet algorithm is used only as a
reference.

\begin{figure}
\includegraphics[width=\columnwidth]{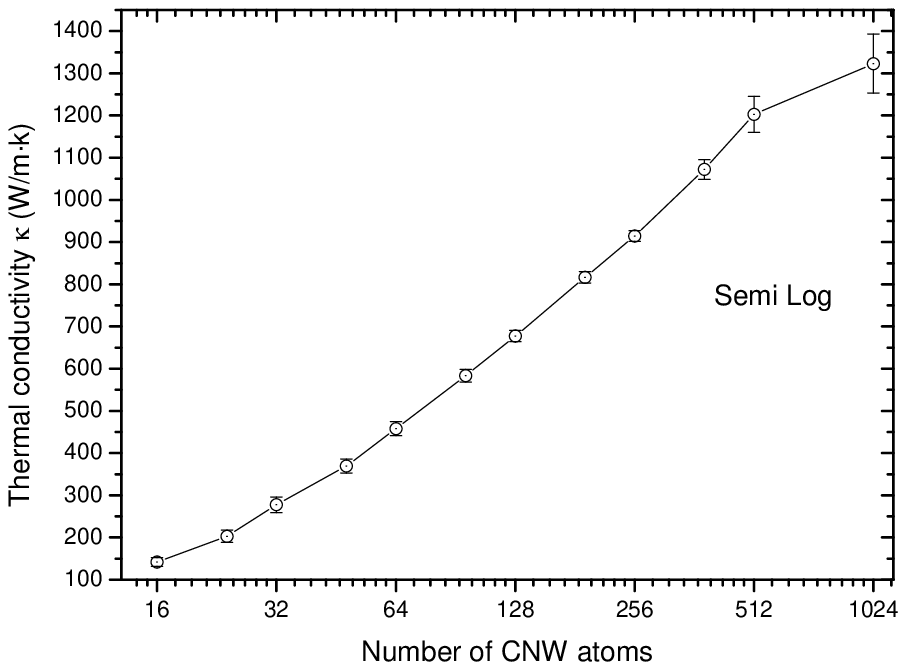}
\label{fig2a}
\includegraphics[width=\columnwidth]{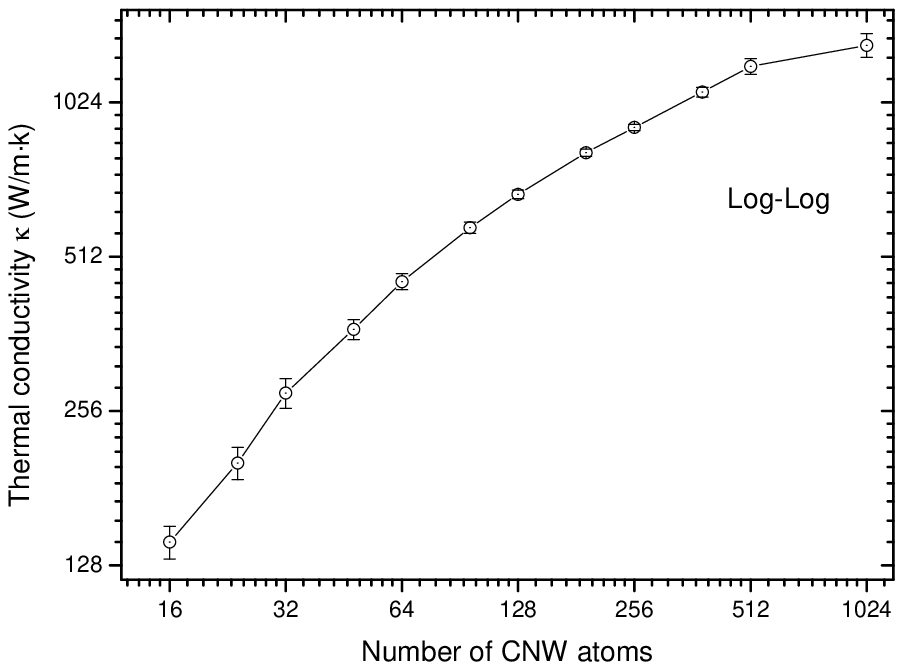}
\label{fig2b} \caption{Thermal conductivity of the CNW as a
function of its length. $a_0$ = 1.844 \AA. a) Linear-log plot. b)
Log-log plot. The solid lines in a) and b) are used for guiding
eyes.}
\end{figure}

The calculated thermal conductivities of CNWs with different
lengths are shown in Fig. 2, in which two types of scales are
shown, i.e., linear-log and log-log scale. Here we should explain
our definition on the cross section of CNW. The statistical radial
distribution for the motion of wire atoms along the direction
perpendicular to the tube axis is calculated, and obtained result
is given in Fig. 3, which can be fitted by $f\left( r \right) = A
\cdot r \cdot \exp \left[ { - \left( {\textstyle{r \over b}}
\right)^2} \right]$, with $r$ the radial distance to the tube
axis. It is found that the parameter $b$ is equal to 0.151 \AA,
and so, the final cross section area for the heat transport can be
expressed as $4\pi b^2$.

\begin{figure}
\includegraphics[width=\columnwidth]{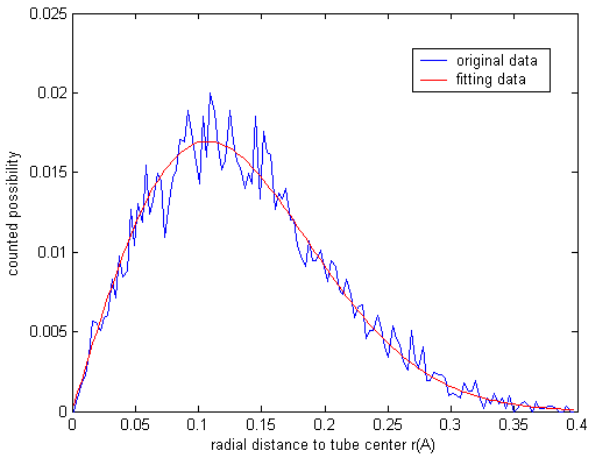}
\caption{\label{fig3}Statistical radial distribution for the
motion of wire atoms perpendicular to the tube axis.}
\end{figure}

Thus in this case, the thermal conductivity $\kappa $ is obtained
to be 142 W/mK$\sim $1323 W/mK, for the chain length $L$ from 3 nm
to 188 nm, which is very high. For comparison, it is interesting
to list the $\kappa $ value of the SWNTs with their lengths in the
same range, which is about 10$^{2}$ W/mK [20], or about
10$^{2}\sim $10$^{3}$ W/mK [21, 22]. So, the thermal conductivity
of CNW is comparable to that of the SWNTs, at least, not much
smaller than that of the SWNT. For example, our calculated thermal
conductivity of the CNW with 512 atoms is 1.2x10$^{3}$ W/mK. But
Zhang et al. [21] got the corresponding thermal conductivity of
nearly 1.6x10$^{3}$ W/mK for the (5, 5) SWNT having 384 layers
(its length is the same as that of the CNW), which is on the same
order as our CNW result.

Now, we would like to ask a question: which type of divergence
behavior does our CNW system follow? Power law or logarithmic law?
From Fig. 2 it is clearly seen that when the system length is
increased, the $\kappa $ does NOT show completely a linear
behavior in both of linear-log and log-log scales, making
difficult to justify which type of divergent behavior, power or
logarithmic law is suitable to the CNW. But, comparing Fig. 2a
with 2b, we could conclude that the $\kappa$ of the CNW prefers
more the logarithmic divergence than the power law, at least for
the middle part of the data from about 32 to 512 atoms, which
demonstrates the divergence behavior of the CNW is different from
that of the SWNT, following the power law. Why logarithmic for the
CNW? We think it is just the transverse motions of the carbon
atoms on the CNW to relax its thermal conductance divergence, and
lead it to deviate from the 1D power law divergence.

\begin{figure}
\includegraphics[width=\columnwidth]{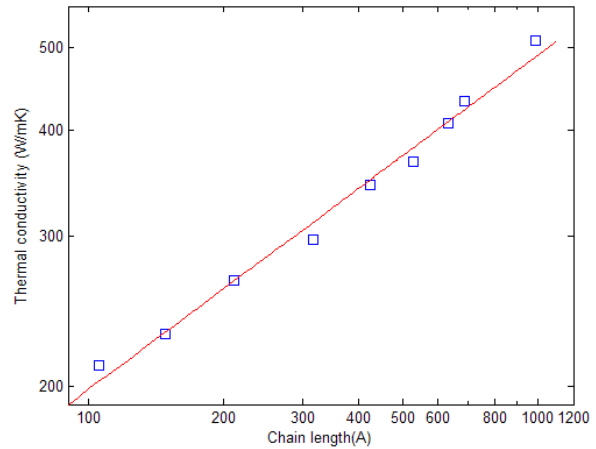}
\caption{\label{fig4}The length dependence of thermal conductivity
of perfect 1D carbon chain model.}
\end{figure}

In order to see more clearly the influence of the transverse
motion of the CNW, we have also studied the thermal conductivity
of perfect 1D carbon chains connected by the Tersoff-Brenner bond
order potential, in which, their transverse motions are not
permitted. The initial equilibrium distance between their
neighboring atoms is set to be 1.65 \AA, and their cross sections
are determined as follows. As well known, the cutoff distance in
the Tersoff-Brenner bond order potential is about 2.0 \AA, which
can be approximately considered as an interaction range between
two nearby carbon chains. So, the cross section of a perfect 1D
carbon chain can be roughly estimated to be about 3.14
{\AA}$^{2}$. The length dependence of the thermal conductivity is
shown in Fig. 4. It is seen from Fig. 4 that the $\kappa$ diverges
with chain length as, $\kappa \propto L^\beta $, with $\beta
\approx 0.39\pm 0.02$, and its $\kappa $ is about 212 W/m K$\sim
$511 W/m K, which is comparable with the result of CNW.

Comparison between the thermal conductivities of both 1D carbon
chain and the quasi-1D CNW clearly show that it is indeed the
transverse motions of the carbon atoms on the CNW to cause its
logarithmic thermal conductance divergence. We should emphasize
that in real systems, the divergence of thermal conductivity will
not be as simple as that found in the theoretical models such as
FPU model, which probably rests with the aspect ratio of the
system.

\begin{figure}
\includegraphics[width=\columnwidth]{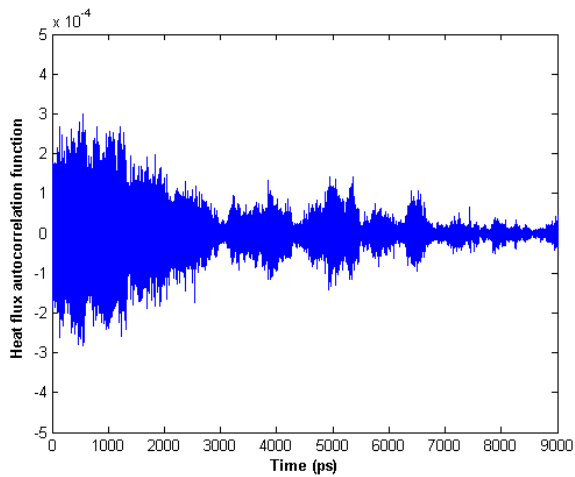}
\caption{\label{fig5}Heat flux autocorrelation function of CNW
with 64 particles.}
\end{figure}

The heat flux autocorrelation function in the case of N=64 is also
calculated and shown in Fig. 5, from which it is seen that after a
very slow decay in about 9000 ps, the heat flux autocorrelation
function decreases to a small value. The similar phenomenon has
been observed by Yao \textit{et al.} [22], which can be understood
by the fact that after a long enough evolution, the final state
has no relationship with the initial state.

Finally, we will check the validity of ensemble average in this
low dimensional system. First, we compare those evolutions
starting from the different initial conditions. The obtained
results are shown in Fig. 6. Here the low or high initial
temperature means the initial temperature of every atom on the
chain is set to the lower or higher boundary temperature,
respectively. And the linear initial temperature means the initial
temperature of every atom is chosen based upon a linear
temperature distribution between the two boundaries. All the
distributions are calculated after a time interval $ \approx
2\times 10^5$.

\begin{figure*}
\includegraphics[width=0.9\columnwidth]{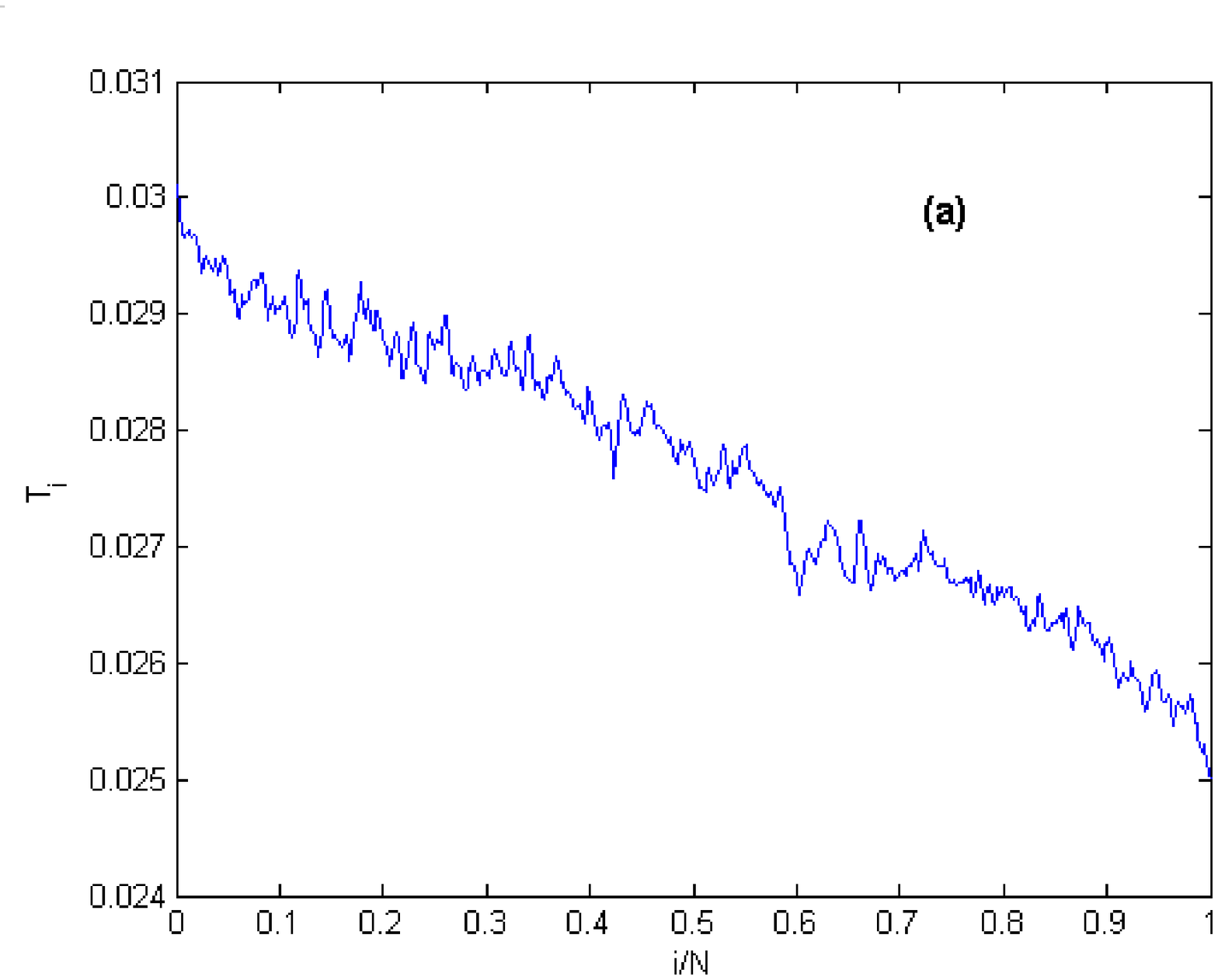}
\label{fig6a}
\includegraphics[width=0.9\columnwidth]{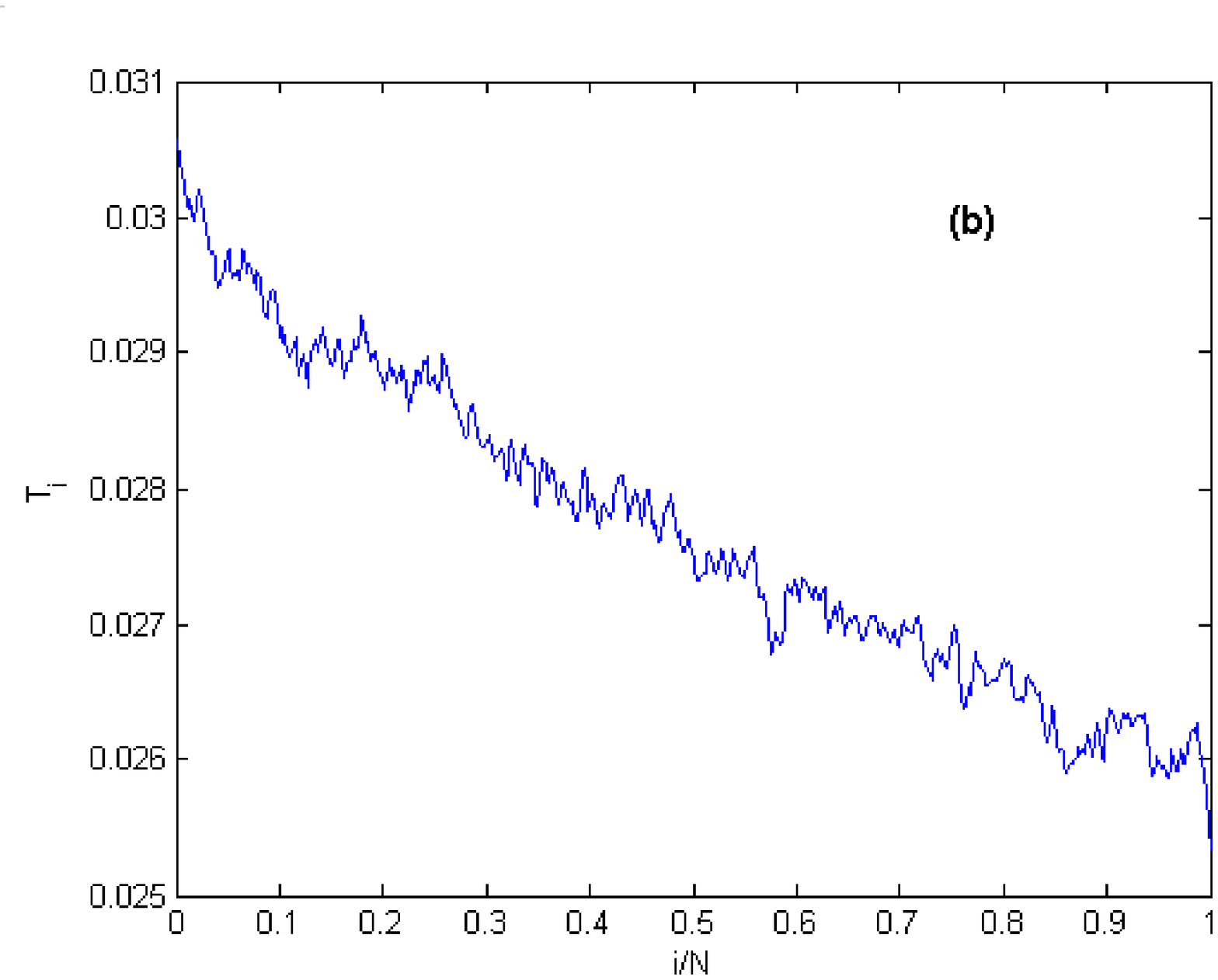}
\label{fig6b}
\includegraphics[width=0.9\columnwidth]{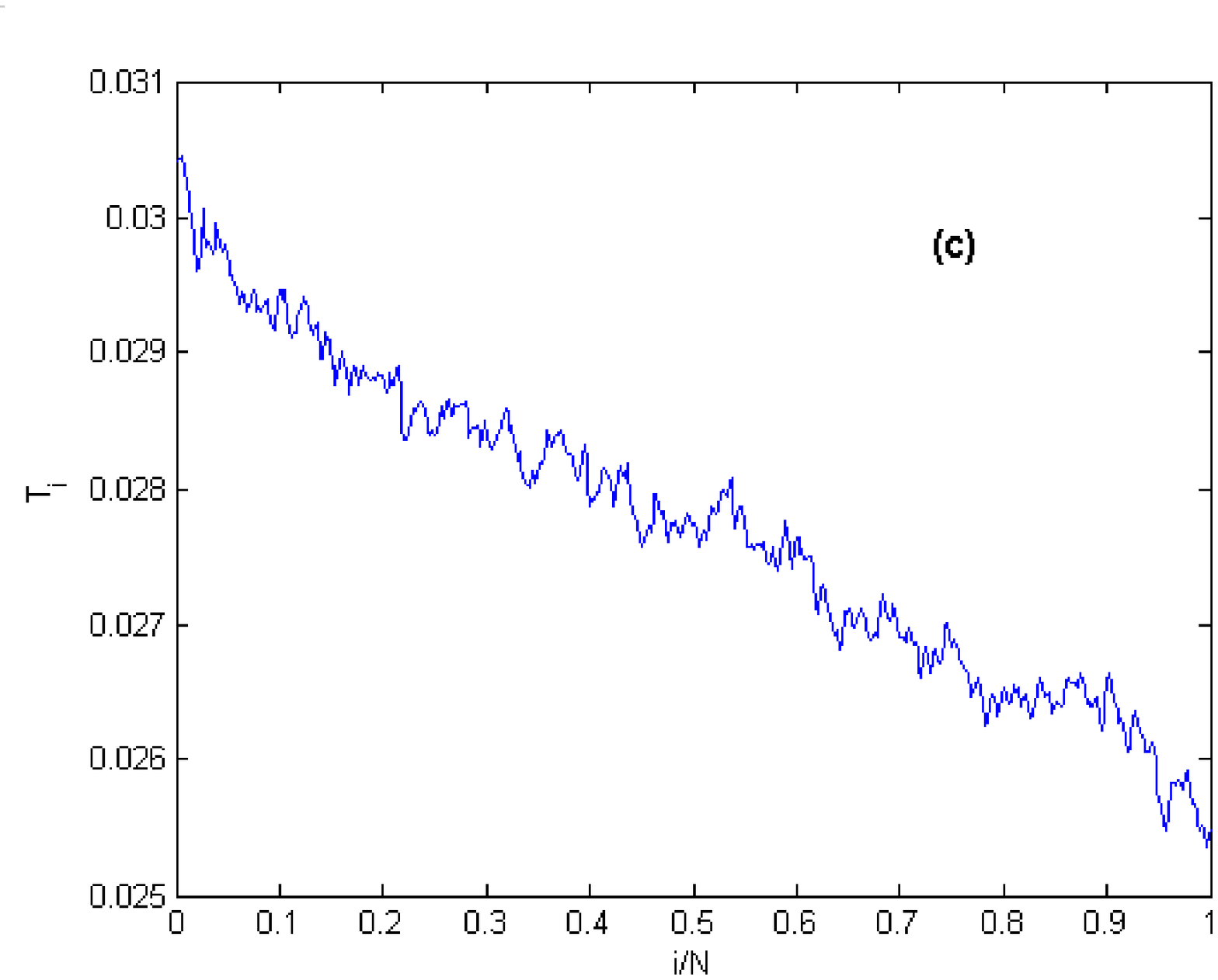}
\label{fig6c}
\includegraphics[width=0.9\columnwidth]{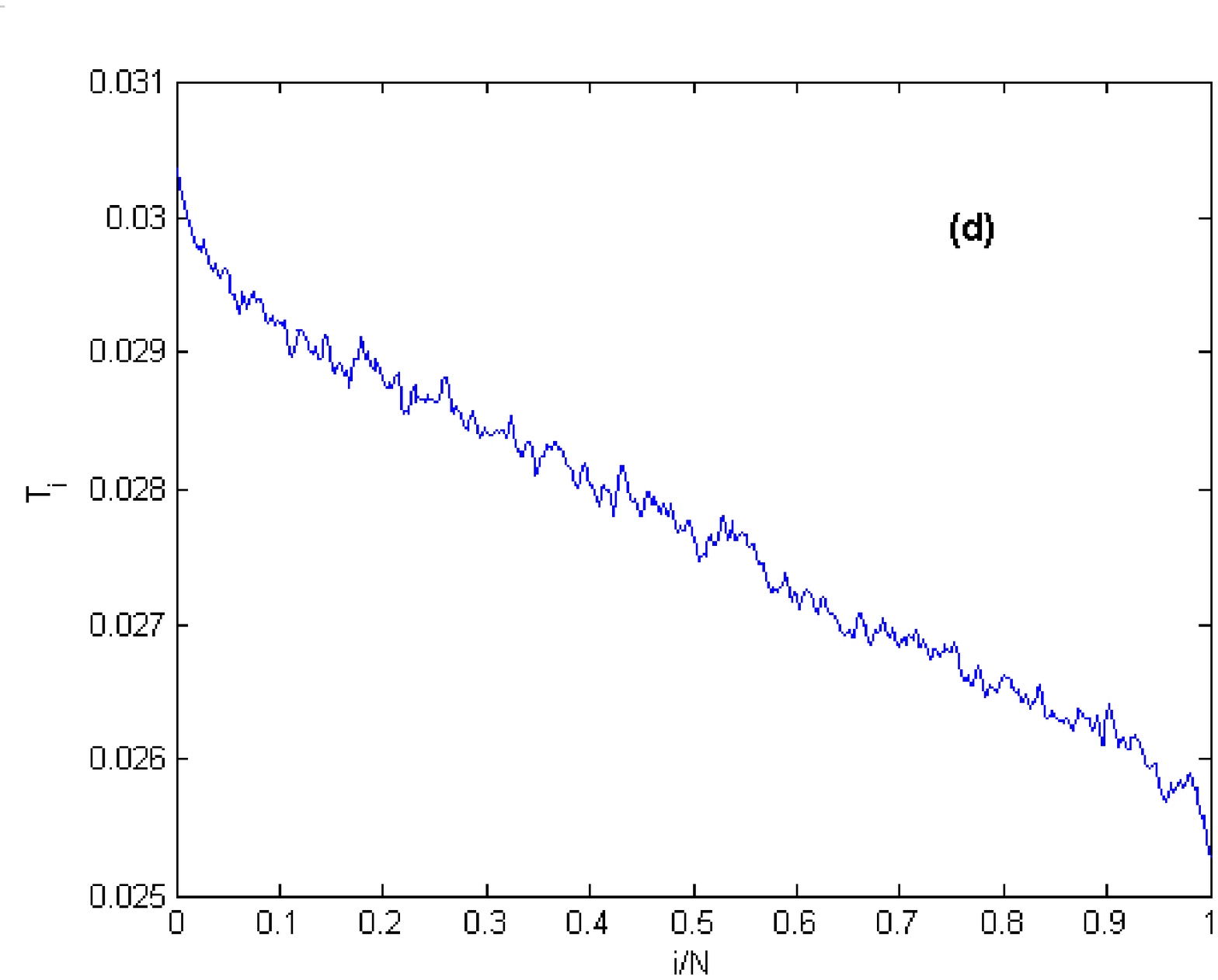}
\label{fig6d} \caption{The final temperature distributions of 512
particles evolved from different initial temperature profile. a).
low initial temperature. b). linear initial temperature. c). high
initial temperature. d). average of a), b) and c).}
\end{figure*}

At the first sight, four figures in Fig. 6 seem to be the same. In
fact there are only little differences between them, which means
our numerical results are reasonable, being independent of the
initial conditions. However, it is seen from Fig. 6 that Fig. 6(d)
is the smoothest one, which means an average over other three can
give a more reliable result, just like averaging over a longer
time interval. Thus, we can improve our calculation efficiency by
the following method: very different initial states are chosen
first, from which the system evolves, and after a period of
evolution time, an average over the different system evolutions
starting from the different initial conditions can be made. This
method can be considered as a high efficient parallel algorithm,
by which, the highest acceleration coefficient can be gained.

\section{Conclusion}

In this paper, the heat conduction of a finite length carbon chain
inserted into a (5, 5) SWNT has been studied by using the
nonequilibrium molecular dynamics method, in which both
longitudinal and transverse motions of the chain atoms are
permitted. The interaction between chain atoms and nanotubes has
been simulated by a continuous model for the nanotube. It is found
that heat conduction of CNW does not obey the Fourier's law, and
its thermal conductivity $\kappa $ logarithmicly diverges with CNW
length $L$ as $\kappa \sim \log \left( L \right)$.

\begin{acknowledgments}
We acknowledge support from the Natural Science Foundation of
China under Grants No. 90103038 and No. A040108.
\end{acknowledgments}

\end{document}